\documentclass[conference]{IEEEtran}
\IEEEoverridecommandlockouts
\usepackage{cite}
\usepackage{amsmath,amssymb,amsfonts}
\usepackage{algorithmic}
\usepackage{graphicx}
\usepackage{textcomp}
\usepackage{xcolor}
\def\BibTeX{{\rm B\kern-.05em{\sc i\kern-.025em b}\kern-.08em
    T\kern-.1667em\lower.7ex\hbox{E}\kern-.125emX}}

\usepackage{subcaption}
\usepackage{tikz}
\usetikzlibrary{quantikz2}
\usepackage{subcaption}
\usepackage{comment}
\usepackage{booktabs}

\begin{document}

\title{Evaluating Variational Quantum Circuit Architectures for Distributed Quantum Computing
\thanks{
© 2025 IEEE.  Personal use of this material is permitted.  Permission from IEEE must be obtained for all other uses, in any current or future media, including reprinting/republishing this material for advertising or promotional purposes, creating new collective works, for resale or redistribution to servers or lists, or reuse of any copyrighted component of this work in other works.

Sponsored in part by the Bavarian Ministry of Economic Affairs, Regional Development and Energy as part of the 6GQT project, as well as the German Federal Ministry of Research, Technology and Space under the funding code 01MQ22008A. The sole responsibility for the report's contents lies with the authors.
}
}

\author{\IEEEauthorblockN{Leo Sünkel}
\IEEEauthorblockA{\textit{Institute for Informatics} \\
\textit{LMU Munich}\\
Munich, Germany \\
leo.suenkel@ifi.lmu.de}
\and
\IEEEauthorblockN{Jonas Stein}
\IEEEauthorblockA{\textit{Institute for Informatics} \\
\textit{LMU Munich}\\
Munich, Germany}
\and
\IEEEauthorblockN{Jonas Nüßlein}
\IEEEauthorblockA{\textit{Institute for Informatics} \\
\textit{LMU Munich}\\
Munich, Germany}
\and
\IEEEauthorblockN{Tobias Rohe}
\IEEEauthorblockA{\textit{Institute for Informatics} \\
\textit{LMU Munich}\\
Munich, Germany}
\and
\IEEEauthorblockN{Claudia Linnhoff-Popien}
\IEEEauthorblockA{\textit{Institute for Informatics} \\
\textit{LMU Munich}\\
Munich, Germany}
}

\maketitle

\begin{abstract}
Scaling quantum computers, i.e., quantum processing units (QPUs) to enable the execution of large quantum circuits is a major challenge, especially for applications that should provide a quantum advantage over classical algorithms. One approach to scale  QPUs is to connect multiple machines through quantum and classical channels to form clusters or even quantum networks. Using this paradigm, several smaller QPUs can collectively execute circuits that each would not be able on its own. However, communication between QPUs is costly as it requires generating and maintaining entanglement, and hence it should be used wisely. In this paper, we evaluate the architectures, and in particular the entanglement patterns, of variational quantum circuits in a distributed quantum computing (DQC) setting. That is, using Qiskit, we simulate the execution of an eight qubit circuit using four QPUs each with two computational and two communication qubits where non-local CX-gates are performed using the remote-CX protocol. We compare the performance of various circuits on a binary classification task where training is executed under ideal and testing under noisy conditions. The study provides initial results on suitable VQC architectures for the DQC paradigm, and indicates that a standard VQC baseline is not always the best choice, and alternative architectures that use entanglement between QPUs sparingly deliver better results under noise.
\end{abstract}


\section{Introduction}
The field of quantum computing (QC) has been rapidly advancing in recent years; in particular, quantum machine learning (QML) \cite{schuld2015introduction,biamonte2017quantum,cerezo2022challenges} applications are being applied and investigated in various domains, including medical \cite{maheshwari2022quantum,wei2023quantum,ullah2024quantum} and finance \cite{orus2019quantum,pistoia2021quantum}. To fully take advantage of the quantum realm, the field must solve the scalability problem, among other challenges. While in the current NISQ-era \cite{preskill2018quantum} QPUs contain a few hundred or even thousand qubits, running large-scale algorithms with error correction requires vastly more. So far the emphasis has mostly been on scaling up single chips, i.e., QPUs, however, the notion of connecting multiple QPUs through quantum communication channels is gaining attention. Multiple QPUs can be directly connected on-site, forming a QC cluster. Alternatively, they could also be connected over large distances forming quantum communication networks and even a possible quantum internet \cite{kimble2008quantum}. These connected devices enable distributed quantum computing (DQC) \cite{caleffi2024distributed} where QPUs collectively execute quantum circuits. More importantly, this allows for a range of small QPUs to run circuits that would not be possible for a QPU to do on its own, providing a pathway to solving the scalability problem. However, to enable large-scale quantum communication requires entanglement to be generated and maintained over large distances, thereby introducing new problems \cite{cacciapuoti2019quantum}. 

Quantum repeaters \cite{briegel1998quantum,van2013designing,azuma2023quantum} that perform entanglement swapping and purification are required, as well as compilers for DQC \cite{ferrari2021compiler}. As quantum entanglement is the central resource in such a network, its use may add another dimension when designing algorithms or circuits. This leads to the main topic of this work, namely the role of entanglement in the form of CX-gates in variational quantum circuits (VQCs). The main focus of VQC design often is the trainability and other performance metrics such as accuracy in a classification task. 

In this work, we evaluate circuit architectures as well as their suitability for the DQC paradigm. More specifically, we train various circuits with different entanglement patterns and frequency. We simulate the distributed execution in a single circuit by creating a large circuit where different qubits represent QPUs consisting of computational and communication qubits. We then add remote-CX between qubits that are not assigned to the same QPU. We simulate these circuits under noise and show that the common VQC architectures are less noise-robust as circuits designed that minimize the remote operations between QPUs. However, the experiments also reveal that a certain amount of entanglement between QPUs is required to achieve acceptable results.

This paper is structured as follows. We recap the necessary background of variational quantum algorithms and DQC in Sec. \ref{sec:background}. We discuss related work in Sec. \ref{sec:related_work} and introduce our approach and evaluated circuit architectures in Sec. \ref{sec:approach}. The experimental setup and results are given in Sec. \ref{sec:experimental_setup} and Sec. \ref{sec:results} respectively. We conclude in Sec. \ref{sec:conclusion}

\section{Background}\label{sec:background}
We begin this section with a brief recapitulation of variational quantum algorithms followed by a short overview of the relevant quantum communication protocols used in the circuits evaluated as part of this work. 

\subsection{Variational Quantum Algorithms}
Variational quantum algorithms (VQAs) \cite{cerezo2021variational,mitarai2018quantum,benedetti2019parameterized} define a broader class of hybrid algorithms consisting of quantum and classical components. In classification tasks, the quantum part is a VQC that consists of data encoding and parameterized layers made up of entangling and parameterized rotation gates. The classical component is an optimization algorithm (e.g. gradient descent, SPSA or COBYLA) that aims to adjust the parameters (i.e., the rotation angles) in order to minimize a given cost function: $\min_\theta C(\theta)$.
In the following, we present an approach relevant for this work; however, other approaches and techniques could also be used. For classification problems, the expectation values of a subset of qubits can be used for class prediction and utilized in the loss, i.e., cost function. More formally, the expectation value of a Pauli observable $\sigma_z$ is defined as:

\begin{equation}
    \mathbb{E}(\sigma_z) = \langle\psi|\sigma_z|\psi\rangle
\end{equation}

The expectation value of the Pauli observable for a qubit $i$ is defined as:

\begin{equation}
    \mathbb{E}(\sigma_z^{(i)}) = \langle0^{\otimes n}|U^\dagger (x, \theta) \sigma_z^{(i)} U(x, \theta)|0^{\otimes n}\rangle
\end{equation}

\noindent where $U$ is a unitary.
For $K$-class classification, the following quantum model in which $K$ qubits are measured can be used:

\begin{equation}
    f(x, \theta) = (\mathbb{E}(\sigma_z^{(0)}), \dots, \mathbb{E}(\sigma_z^{(K-1)}))
\end{equation}

\noindent i.e., the VQC is given a data point $x$ and parameters $\theta$, and returns a vector containing the expectation values. The extracted expectation values of the observable $\sigma_z$ are the output of the model. To obtain class probabilities, the softmax function can be applied, which is defined as:

\begin{equation}
    p(z)_j = \frac{e^{z_j}}{\sum_l e^{z_l}}
\end{equation}

\noindent where $z$ is a vector. 
The cross-entropy loss function can be used which is defined as follows:

\begin{equation}
   \mathcal{L}(y, p) =  - \frac{1}{N} \sum_{i=0}^{N-1} \sum_{k=0}^{K-1}  y_{i,k} \textnormal{log} p_{i,k}
\end{equation}

Thus, the cost function becomes

\begin{equation}\label{eqc:cost}
    C(\theta) = \mathcal{L}(y, p(f(X, \theta))),
\end{equation}

\noindent where $\theta$ are the parameters, $y$ the true labels, and $X$ the dataset.
The objective is then:

\begin{equation}
    \min_\theta C(\theta)
\end{equation}

The model (i.e., VQC) is then trained iteratively where the circuit is executed on a quantum computer (or simulator) while a classical computer processes the measurement results (e.g., applying softmax), which can then be used as class probabilities in the optimization algorithm to adjust the parameters accordingly.

\subsection{Distributed Quantum Computing}
QPUs connected via classical and quantum communication channels can be utilized to perform computations as if they were a single unit in a paradigm known as DQC. QPUs can be arranged in a cluster in close proximity or over large distances in quantum networks; however, the latter approach poses significant challenges in both hardware and software. To enable large-distance quantum communication over networks, so-called quantum repeaters performing entanglement swapping and purification can be placed throughout the network. With the entanglement swapping protocol, two qubits can be entangled without directly interacting with each other. The protocol requires two Bell pairs. With a purification protocol, $n$ weakly entangled qubits can be used to increase the entanglement of $k$ qubits, where $k < n$. These are high-level requirements for quantum communication networks.
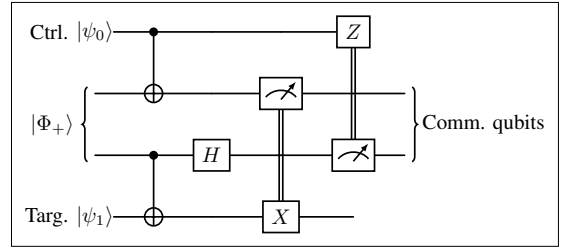
\begin{figure}
    \centering
    \scalebox{0.8}{
        \boxed{
            \begin{quantikz}
                    \lstick[1]{Ctrl.} {\ket{\psi_0}} & \ctrl{1}  &&& \gate{Z} \\
                    \lstick[2]{\ket{\Phi_+}} & \targ{} &  & \meter{} \wire[d][2]{c} && \rstick[2]{Comm. qubits}   \\
                    & \ctrl{1} & \gate{H}  && \meter{} \wire[u][2]{c}&  \\
                    \lstick[1]{Targ.} {\ket{\psi_1}} & \targ{} & & \gate{X} &    
                \end{quantikz}
        }
        }
    \caption{The remote CX protocol. The top qubit $\ket{\psi_0}$ acts as the control and the bottom qubit $\ket{\psi_1}$ as the target. Two entangled communication qubits are required for this protocol. \cite{caleffi2024distributed}}
    \label{fig:remote_cx_protocol}
\end{figure}
In DQC, qubits are allocated to different QPUs, ideally in a way such that the number of remote operations, i.e., multi-qubit gates where qubits reside in different QPUs, is minimized. However, when remote operations do need to be performed, qubits can be teleported such that they are located in the same QPU and the gate can subsequently be executed locally. Alternatively, a remote-CX can be performed in which the qubits remain in their QPUs and are not moved. Both strategies have advantages depending on the specific circuit; however, this is its own optimization problem and not part of this work; in this work we solely use the remote-CX protocol, which is depicted in Fig. \ref{fig:remote_cx_protocol}. The protocol requires a shared Bell pair, where each QPU contains one qubit of said pair.

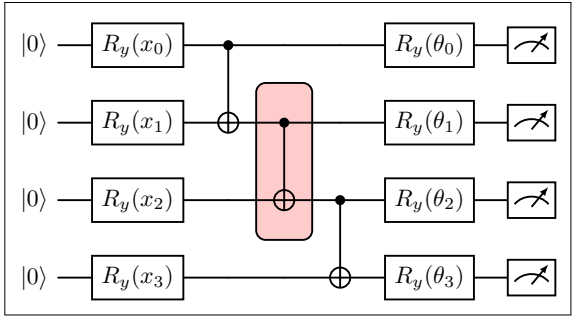
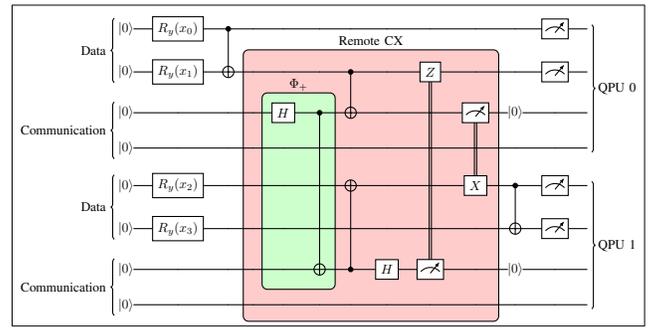
\begin{figure*}[tb]
    \centering
    \begin{subfigure}[t]{0.45\textwidth}
        \centering
        \scalebox{0.9}{
        \boxed{
        \begin{quantikz}
            \lstick{\ket{0}} & \gate{R_y{(x_0)}} & \ctrl{1} & & & \gate{R_y{(\theta_0)}} & \meter{} \\
            \lstick{\ket{0}} & \gate{R_y{(x_1)}} & \targ{} & \ctrl{1} \gategroup[2, steps=1, style={rounded corners, fill=red!20}, background]{} & & \gate{R_y{(\theta_1)}} & \meter{} \\
            \lstick{\ket{0}} & \gate{R_y{(x_2)}} & & \targ{} & \ctrl{1} & \gate{R_y{(\theta_2)}} & \meter{} \\
            \lstick{\ket{0}} & \gate{R_y{(x_3)}} & & & \targ{} & \gate{R_y{(\theta_3)}} & \meter{}    
        \end{quantikz}
        }
        }
        \caption{Example VQC with one remote-CX (red).}
        \label{fig:example_vqc_monolithic}
    \end{subfigure}
    \hfill
    \begin{subfigure}[t]{0.45\textwidth}
        \centering
        \scalebox{0.5}{
        \boxed{
            \begin{quantikz}
               \lstick[2]{Data} \ket{0} & \gate{R_y{(x_0)}} & \ctrl{1} &&&&&&&&& \meter{} &  \rstick[4]{QPU 0} \\
                \ket{0} & \gate{R_y{(x_1)}} & \targ{} &\gategroup[7, steps=7, style={rounded corners, fill=red!20}, background, label style={label position=above}]{Remote CX} &&& \ctrl{1} && \gate{Z} &&&\meter{} & \\
                \lstick[2]{Communication} \ket{0} &&&& \gate{H}\gategroup[5, steps=2, style={rounded corners,fill=green!20}, background, label style={label position=above}]{$\Phi_+$} & \ctrl{4} & \targ{} &&& \meter{} \wire[d][2]{c} & \ket{0} && \\
                \ket{0} &&&&&&&&&&&& \\
                \lstick[2]{Data} \ket{0} & \gate{R_y{(x_2)}} &&&&& \targ{} &&& \gate{X} & \ctrl{1} & \meter{} & \rstick[4]{QPU 1} \\
                \ket{0} & \gate{R_y{(x_3)}} &&&&&&&&& \targ{} & \meter{} & \\
                \lstick[2]{Communication} \ket{0} &&&&& \targ{} & \ctrl{-2} & \gate{H} & \meter{} \wire[u][5]{c} && \ket{0} && \\
                \ket{0} &&&&&&&&&&&&
            \end{quantikz}
        }
        }
        \caption{Example circuit containing two QPUs each with two data and two communication qubits. The protocol to execute the remote CX gate is depicted in the red box.}
        \label{fig:example_distributed_vqc}
    \end{subfigure}
    \caption{Example of a regular VQC and how it is internally simulated as a distributed circuit. For each remote CX, the protocol is inserted into the circuit an the respective data (i.e., computational) and communication qubits.}
    \label{fig:example_qc_with_remote_cx}
\end{figure*}

\subsection{Problem Statement}
On the one hand, the entanglement layer is a crucial aspect in a VQC; in DQC, however, remote operations between QPUs should be used sparingly to save valuable resources. In QML literature, the question of necessary entanglement has been addressed; however, the focus lies mainly on performance and trainability and not suitability for execution in a DQC environment. In DQC, one would want to retain as much entanglement necessary to achieve an acceptable performance in the usual QML metrics while simultaneously reducing remote operations between QPUs. To this end, we evaluate different VQC architectures with different entangling layers that should be executed using the DQC paradigm. We define a 4 node network where each node has two data and two communication qubits. Moreover, for each remote-CX operation, the protocol is inserted into the circuit and then simulated under noise. Note that is it not a full quantum network simulator, it is rather used to evaluate the influence of remote operations. We discuss our approach in detail in Sec. \ref{sec:approach}. The aim of this study thus is twofold: (1) we evaluate how different entanglement patterns affect the training of the VQC and (2) how robust these circuit architectures are under noise when converted to the distributed version of the circuit.

\section{Related Work}\label{sec:related_work}
Executing variational quantum eigensolvers (VQEs) in a DQC-setting has been investigated in \cite{diadamo2021distributed,khait2023variational} while distributed training of VQAs is discussed in \cite{du2022distributed}. DQC for chemistry is discussed in \cite{jones2024distributed}, and distributed QML in \cite{pira2023invitation,neumann2022distributed}. Algorithms for DQC were also discussed in \cite{parekh2021quantum}.  An overview of VQAs is given in \cite{cerezo2021variational}, VQC as classifiers is discussed in \cite{schuld2020circuit}. The expressibility and entangling capability of VQCs is investigated in \cite{sim2019expressibility,hubregtsen2021evaluation} while the influence of entanglement on VQC in \cite{rohe2024questionable}. Designing the architecture of quantum circuits has been gaining attention in recent years, in particular methods that try to automate this process. In \cite{du2022quantum} the authors investigate the search for architectures for VQCs, and in \cite{ding2022evolutionary} the authors apply an evolutionary algorithm to search for parameterized circuits while the distributed architecture search is discussed in \cite{situ2024distributed}. 

\begin{figure*}[t]
    \centering
    \begin{subfigure}[t]{0.45\textwidth}
    \centering
        \scalebox{0.75}{
            \boxed{
                \begin{quantikz}
                    \lstick{\ket{0}} & \gate{R_y{(x_0)}} & \ctrl{1} \gategroup[4, steps=4, style={rounded corners, fill=green!20}, background]{Single Layer} & & & \gate{R_y{(\theta_0)}} & \meter{} \\
                    \lstick{\ket{0}} & \gate{R_y{(x_1)}} & \targ{} & \ctrl{1} & & \gate{R_y{(\theta_1)}} & \meter{} \\
                    \lstick{\ket{0}} & \gate{R_y{(x_2)}} & & \targ{} & \ctrl{1} & \gate{R_y{(\theta_2)}} &  \\
                    \lstick{\ket{0}} & \gate{R_y{(x_3)}} & & & \targ{} & \gate{R_y{(\theta_3)}} &     
                \end{quantikz}
            }
        }
    \caption{Baseline VQC. This architecture is commonly applied in QML tasks.}
    \label{fig:baseline_vqc}
    \end{subfigure}
    \begin{subfigure}[t]{0.45\textwidth}
        \centering
        \scalebox{0.75}{
            \boxed{
                \begin{quantikz}
                    \lstick{\ket{0}} & \gate{R_y{(x_0)}} & \ctrl{1} & \ctrl{2} & \ctrl{3} & & & & \gate{R_y{(\theta_0)}} \gategroup[4, steps=1, style={rounded corners, fill=red!20}, background]{Single Layer}  & \meter{} \\
                    \lstick{\ket{0}} & \gate{R_y{(x_1)}} & \targ{} & & & \ctrl{1} & \ctrl{2} & & \gate{R_y{(\theta_1)}} & \meter{} \\
                    \lstick{\ket{0}} & \gate{R_y{(x_2)}} & & \targ{} & & \targ{} & & \ctrl{1} & \gate{R_y{(\theta_2)}} &  \\
                    \lstick{\ket{0}} & \gate{R_y{(x_3)}} & & & \targ{} & & \targ{} & \targ{} & \gate{R_y{(\theta_3)}} &     
                \end{quantikz}
            }
        }
    \caption{A single CX between every pair of qubits once after feature encoding. We refer to this circuit as the fully entangled circuit.}
     \label{fig:full_entanglement_after_encoding_vqc}
    \end{subfigure}

    \begin{subfigure}[t]{0.45\textwidth}
    \centering
        \scalebox{0.75}{
            \boxed{
                \begin{quantikz}
                    \lstick{\ket{0}} & \gate{R_y{(x_0)}} & \ctrl{1} \gategroup[4, steps=4, style={rounded corners, fill=green!20}, background]{Global Layer} & & & \gate{R_y{(\theta_0)}} & \ctrl{1} \gategroup[4, steps=2, style={rounded corners, fill=red!20}, background]{Local Layer} & \gate{R_y{(\theta_4)}} & \meter{} \\
                    \lstick{\ket{0}} & \gate{R_y{(x_1)}} & \targ{} & \ctrl{1} & & \gate{R_y{(\theta_1)}} & \targ{}  & \gate{R_y{(\theta_5)}} & \meter{} \\
                    \lstick{\ket{0}} & \gate{R_y{(x_2)}} & & \targ{} & \ctrl{1} & \gate{R_y{(\theta_2)}} & \ctrl{1} & \gate{R_y{(\theta_6)}} &  \\
                    \lstick{\ket{0}} & \gate{R_y{(x_3)}} & & & \targ{} & \gate{R_y{(\theta_3)}} & \targ{} & \gate{R_y{(\theta_7)}} &     
                \end{quantikz}
            }
        }
    \caption{A VQC with alternating global and local layers. A global layer contains CX gates between different QPUs whereas a local layer only contains CX between qubits residing in the same QPU.}
    \label{fig:alternating_layers_vqc}
    \end{subfigure}
    
    \begin{subfigure}[t]{1\textwidth}
        \centering
        \scalebox{0.6}{
        \boxed{
            \begin{quantikz}
                \lstick{\ket{0}} & \gate{R_y{(x_0)}} & \gategroup[8, steps=5, style={rounded corners, inner ysep=18pt,fill=green!20}, background]{Global Layer} & \ctrl{1} \gategroup[8, steps=1, style={rounded corners, fill=red!20}, background, label style={label position=above}]{Local CX}   && \gategroup[8, steps=1, style={rounded corners, fill=red!20}, background, label style={label position=below,yshift=-0.5cm}]{Global CX}& \gate{R_y{(\theta_0)}} & && \ctrl{1}\gategroup[8, steps=2, style={rounded corners, inner ysep=18pt, fill=blue!20}, background]{Local Layer}   & \gate{R_y{(\theta_8)}} & \meter{} & \rstick[2]{QPU 0} \\
                \lstick{\ket{0}} & \gate{R_y{(x_1)}} && \targ{} && \ctrl{1} & \gate{R_y{(\theta_1)}} &&& \targ{} & \gate{R_y{(\theta_9)}} & \meter{} & \\
                \lstick{\ket{0}} & \gate{R_y{(x_2)}} && \ctrl{1} && \targ{} & \gate{R_y{(\theta_2)}} &&& \ctrl{1} & \gate{R_y{(\theta_{10})}} & & \rstick[2]{QPU 1} \\
                \lstick{\ket{0}} & \gate{R_y{(x_3)}} & & \targ{} && \ctrl{1} & \gate{R_y{(\theta_3)}} &&& \targ{} & \gate{R_y{(\theta_{11})}} & & \\
                \lstick{\ket{0}} & \gate{R_y{(x_4)}} && \ctrl{1} && \targ{} & \gate{R_y{(\theta_4)}} &&& \ctrl{1} & \gate{R_y{(\theta_{12})}} &  & \rstick[2]{QPU 2} \\
                \lstick{\ket{0}} & \gate{R_y{(x_5)}} && \targ{} && \ctrl{1} & \gate{R_y{(\theta_5)}} &&& \targ{} & \gate{R_y{(\theta_{13})}} &  & \\
                \lstick{\ket{0}} & \gate{R_y{(x_6)}} && \ctrl{1} && \targ{} & \gate{R_y{(\theta_6)}} &&& \ctrl{1} & \gate{R_y{(\theta_{14})}} & & \rstick[2]{QPU 3} \\
                \lstick{\ket{0}} & \gate{R_y{(x_7)}} && \targ{} &&& \gate{R_y{(\theta_7)}} &&& \targ{} & \gate{R_y{(\theta_{15})}} &  &
            \end{quantikz}
        }}
        \caption{A circuit architecture containing global and local layers. Local layers apply CX gates only between qubits within a QPU whereas global layers additionally apply a single CX between consecutive QPUs.}
        \label{fig:gl_architecure}  
    \end{subfigure}
    \caption{The circuit architectures evaluated in this work.}
    \label{fig:vqc_architectures}
\end{figure*}
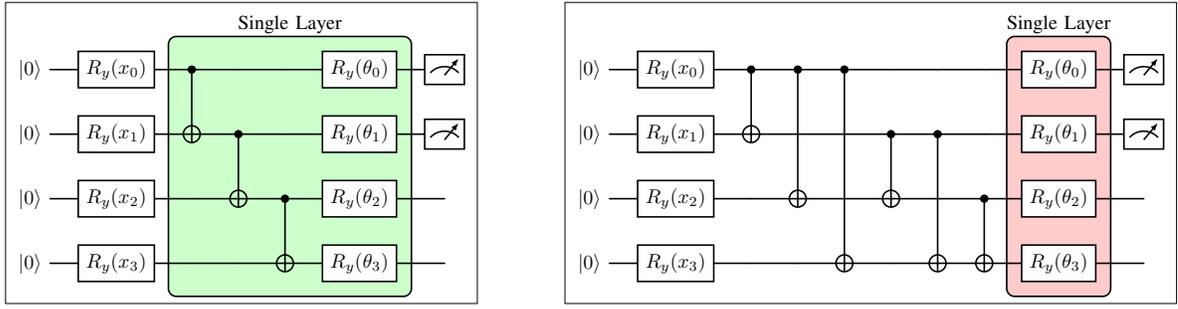
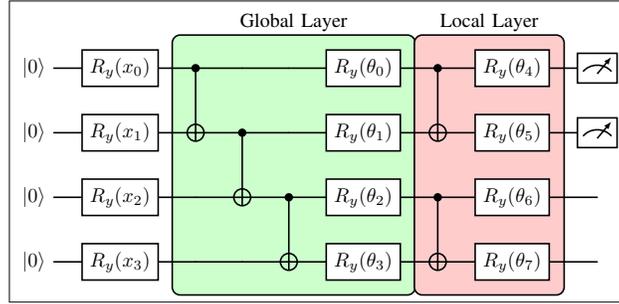
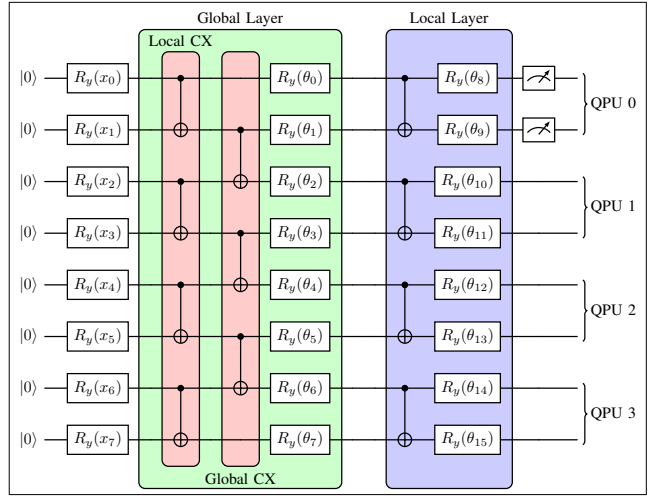

\section{Approach}\label{sec:approach}
We simulate a DQC scenario with 4 nodes, each with two data qubits (or computational qubits) and two communication qubits. A data qubit is used for the computation, i.e., these are the logical qubits used in a VQC; a communication qubit is solely used for communication protocols, i.e., to perform a CX between qubits of different QPUs. However, we do not utilize a quantum network simulator, but rather simulate this architecture using a single Qiskit \cite{qiskit2024} circuit containing 16 qubits in total (8 computational and 8 communication qubits). Remote operations, i.e., remote CX gates, are inserted according to a sequential qubit allocation. An example circuit with two QPUs and one remote operation between them is shown in Fig. \ref{fig:example_qc_with_remote_cx}. A simple VQC is shown in Fig. \ref{fig:example_vqc_monolithic}. Qubits are assigned to QPUs in a sequential manner, that is, the first two qubits are assigned to QPU 0, the next two qubits to QPU 1, and so forth. Therefore, in the example, there is one remote operation between the QPUs, which is marked red. In our approach, this monolithic circuit is converted to a circuit simulating a DQC environment, where each QPU is assigned four consecutive qubits (two for computation and two for communication). For each non-local operation, i.e, a CX-gate between different QPUs, the remote-CX protocol is inserted. The corresponding example is shown in Fig. \ref{fig:example_distributed_vqc}.
This study focuses on different VQC architectures and the effect of remote CX gates; we execute (i.e., simulate) the circuits under noise, but we do not consider network topologies, distance, and other communication protocols (e.g. entanglement swapping or purification) as this is outside the scope of this work. We will return to this fact in the discussion below. 
Our approach works as follows. Circuits are trained with the architectures discussed below without remote operations or noise in a monolithic, i.e. non-distributed fashion. After training, the resulting parameters are used in the DQC setting. That is, remote operations are inserted and the entire DQC circuit is simulated under noise. Thus, we evaluate the training behavior of the different circuit architectures under ideal conditions as well as the testing under noisy conditions with remote operations.
The circuit architectures evaluated in this work are shown in Fig. \ref{fig:vqc_architectures}. A standard VQC used in the literature is shown in Fig. \ref{fig:baseline_vqc} and will be the baseline in the experiments. 
In the architecture depicted in \ref{fig:full_entanglement_after_encoding_vqc}, a CX gate is performed between every qubit once, and each layer contains only parameterized rotations and no entangling gates, we refer to this circuit as the ''fully entangled'' circuit. Fig. \ref{fig:alternating_layers_vqc} shows a circuit with alternating ''global'' and ''local'' layers, where a global layer contains CX gates both within the same QPU as well as different QPUs and a local layer only contains CX gates between qubits located in the same QPU. These layers are applied in an alternating manner. Note that each circuit contains 10 layers in all experiments (cf. Table \ref{tab:experiment_config}). Fig. \ref{fig:gl_architecure} shows a further architecture with two types of layers, i.e., local and global. The local layer in this circuit is identical to the previous one; however, the global layer is slightly different in that it contains local CX gates and CX gate between consecutive QPUs. We refer to this architecture as the ''Alternating Layers 2'' circuit. In the experiments in this work, a global layer is applied every four layers in the ''Alternating Layers 2'' circuit, and each layer is either a local or a global layer. Other variants, e.g. alternating between local and global layers could also be explored in future work, alongside alternative architectures. All circuits contain the same number of trainable parameters (80); the difference between the circuit lies in how and when entangling gates are applied.

\begin{figure*}[tb]
    \centering
    \begin{subfigure}{0.45\textwidth}
        \centering
        \includegraphics[scale=0.42]{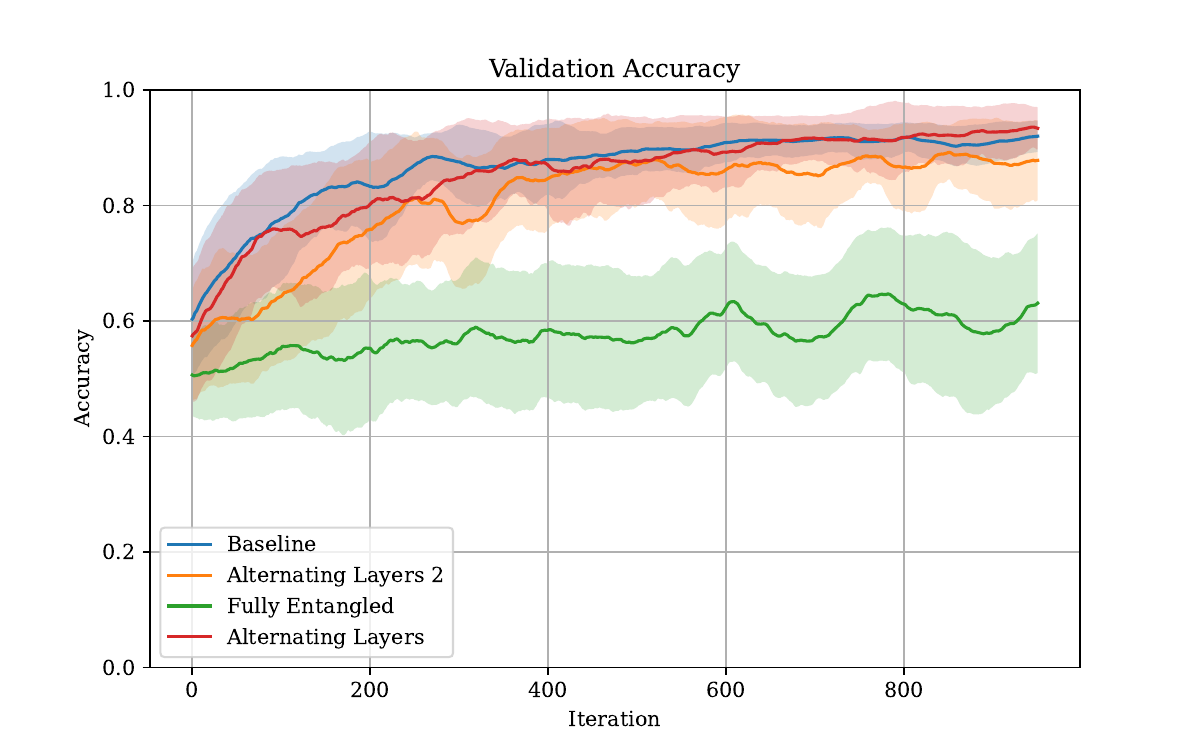}
        \caption{Validation accuracy for dataset 1 during training}
        \label{fig:validation_acc_dataset_1}
    \end{subfigure}
    \hfill
    \begin{subfigure}{0.45\textwidth}
        \centering
        \includegraphics[scale=0.4]{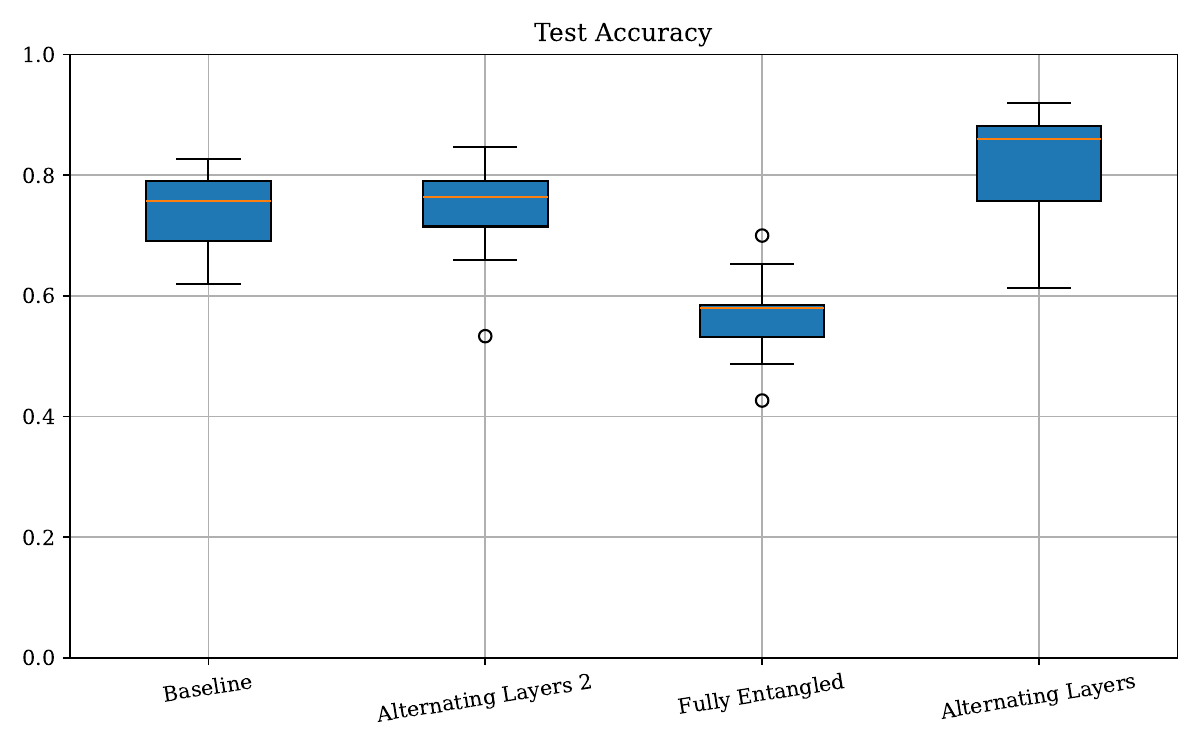}
        \caption{Test accuracy for dataset 1.}
        \label{fig:test_acc_dataset1}
    \end{subfigure}
    \hfill
    \begin{subfigure}{0.45\textwidth}
        \centering
        \includegraphics[scale=0.42]{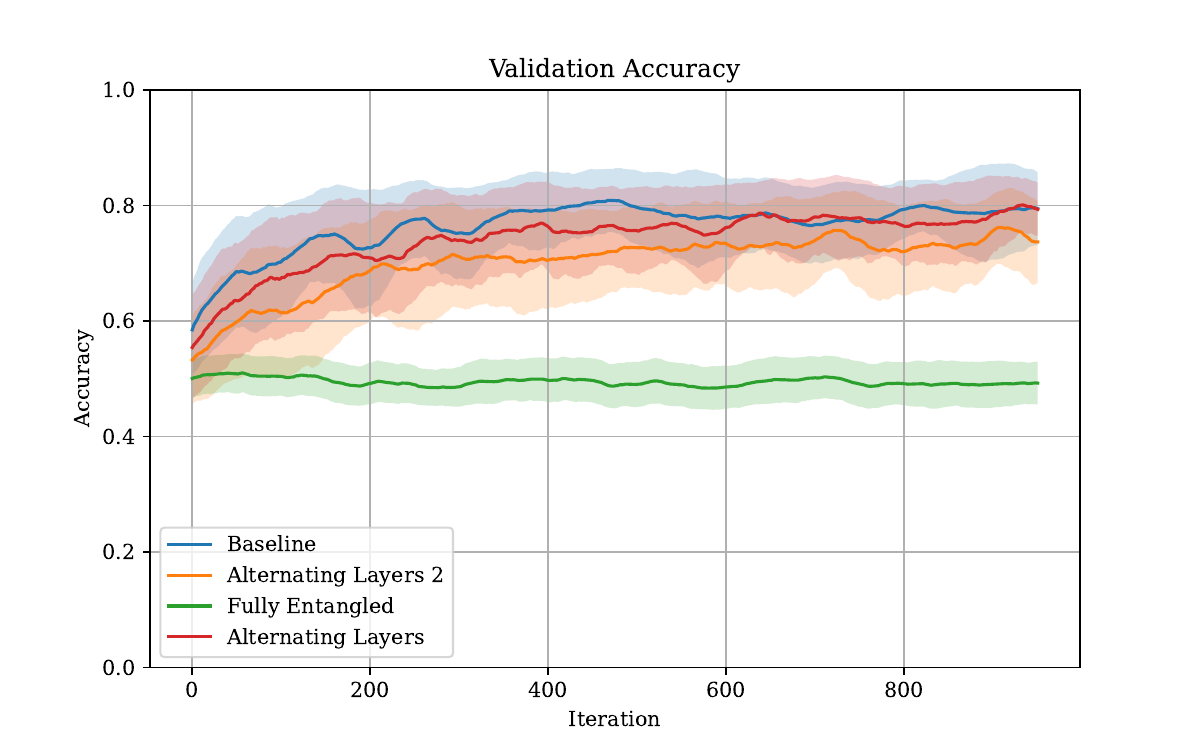}
        \caption{Validation accuracy for dataset 2.}
        \label{fig:validation_acc_dataset_2}
    \end{subfigure}
    \hfill
    \begin{subfigure}{0.45\textwidth}
        \centering
        \includegraphics[scale=0.4]{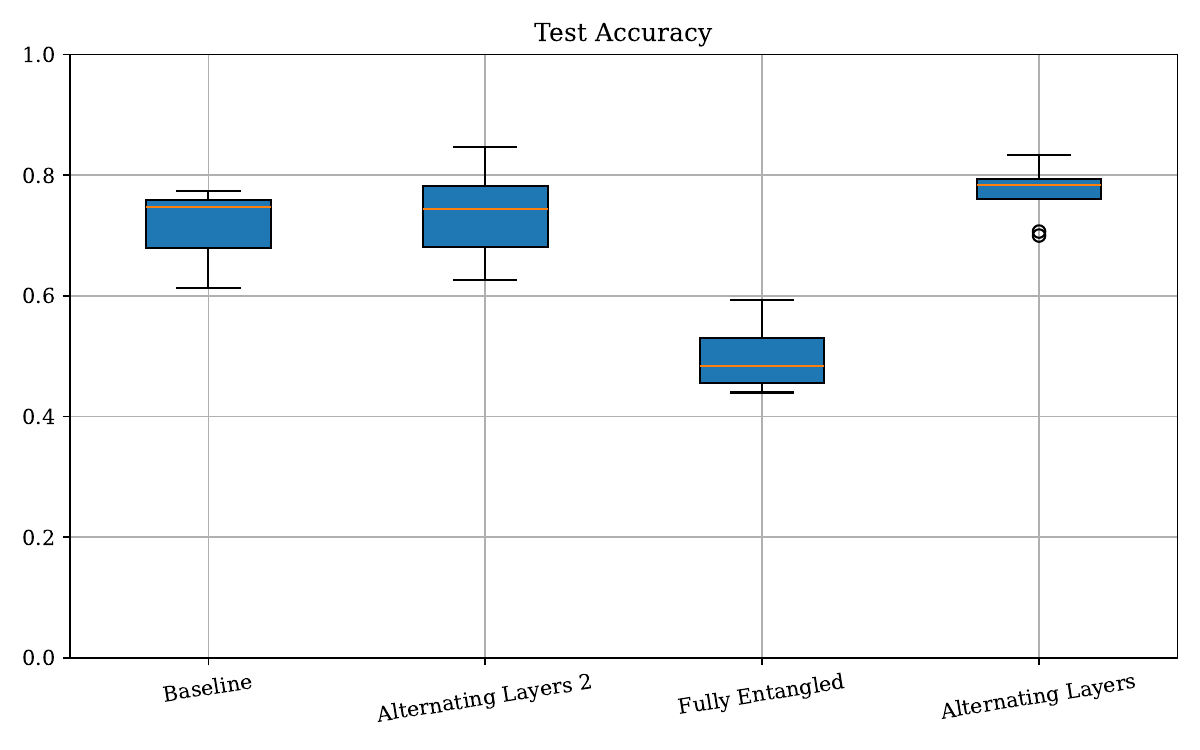}
        \caption{Test accuracy for dataset 2.}
        \label{fig:test_acc_dataset2}
    \end{subfigure}
    \hfill
    \begin{subfigure}{0.45\textwidth}
        \centering
        \includegraphics[scale=0.42]{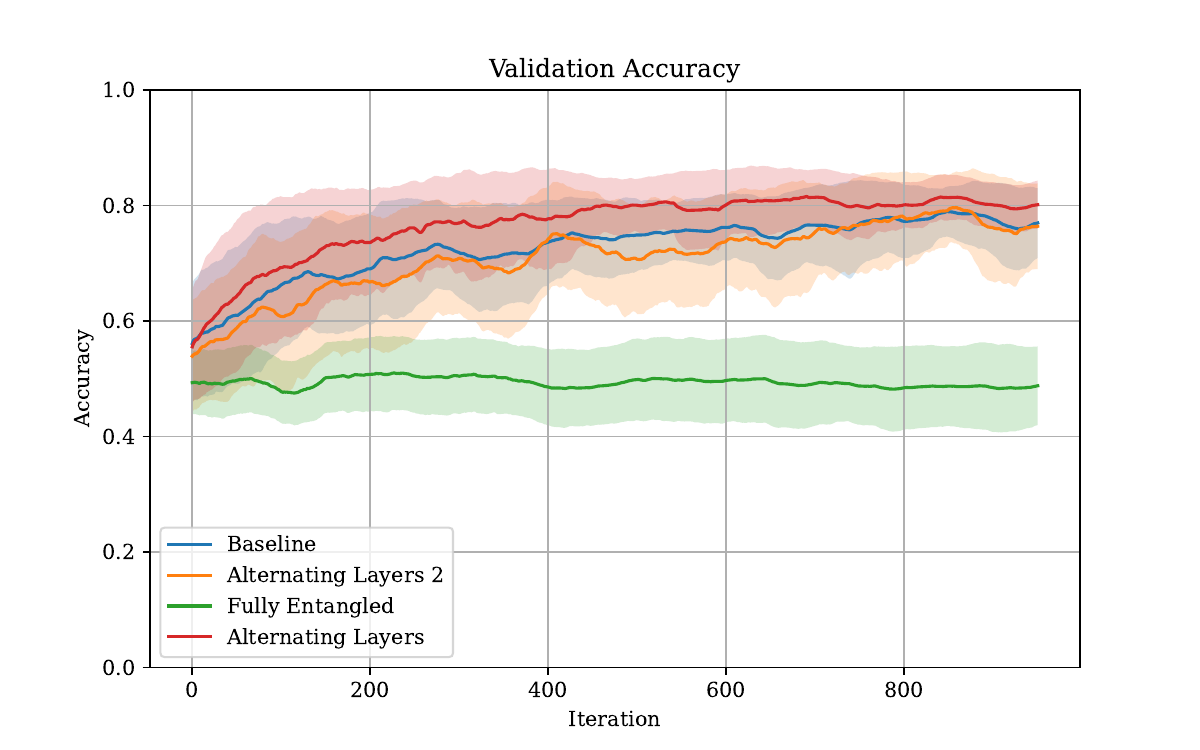}
        \caption{Validation accuracy for dataset 3.}
        \label{fig:validation_acc_dataset_3}
    \end{subfigure}
    \hfill
    \begin{subfigure}{0.45\textwidth}
        \centering
        \includegraphics[scale=0.4]{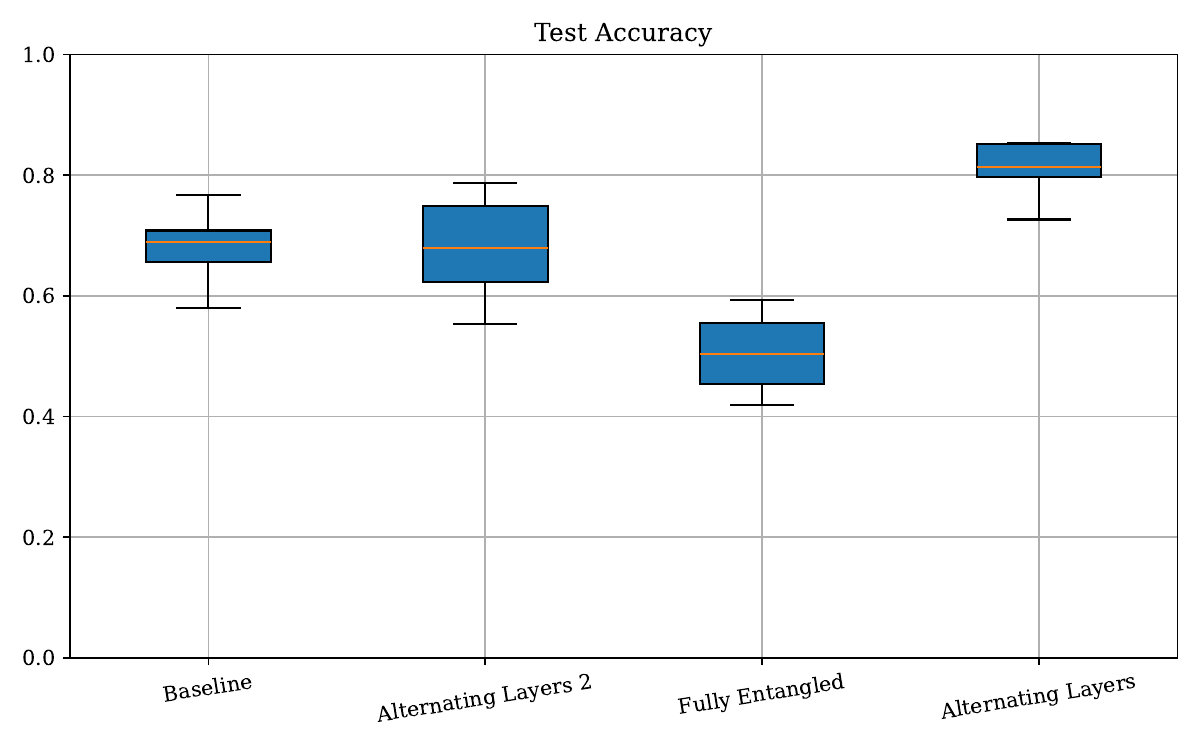}
        \caption{Test accuracy for dataset 3.}
        \label{fig:test_acc_dataset3}
    \end{subfigure}
    \caption{Results of the experiments on all three datasets. Plots on the left are smoothed and show the mean performance over all ten seeds with standard deviation on the validation set.  Plots on the right show results on the test set.}
    \label{fig:results}
\end{figure*}

\begin{figure*}[tb]
    \begin{subfigure}{0.45\textwidth}
        \centering
        \includegraphics[scale=0.4]{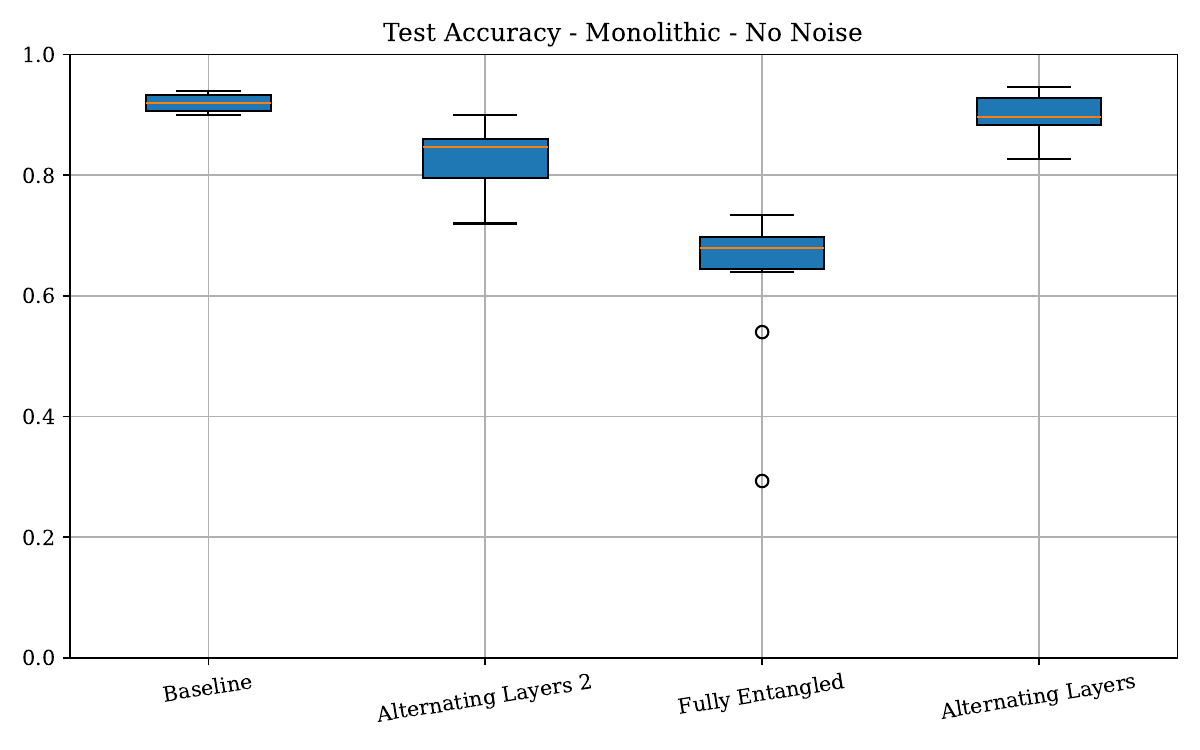}
        \caption{Test accuracy for dataset 1.}
        \label{fig:test_acc_dataset1_monolithic_no_noise}
    \end{subfigure}
    \hfill
    \begin{subfigure}{0.45\textwidth}
        \centering
        \includegraphics[scale=0.42]{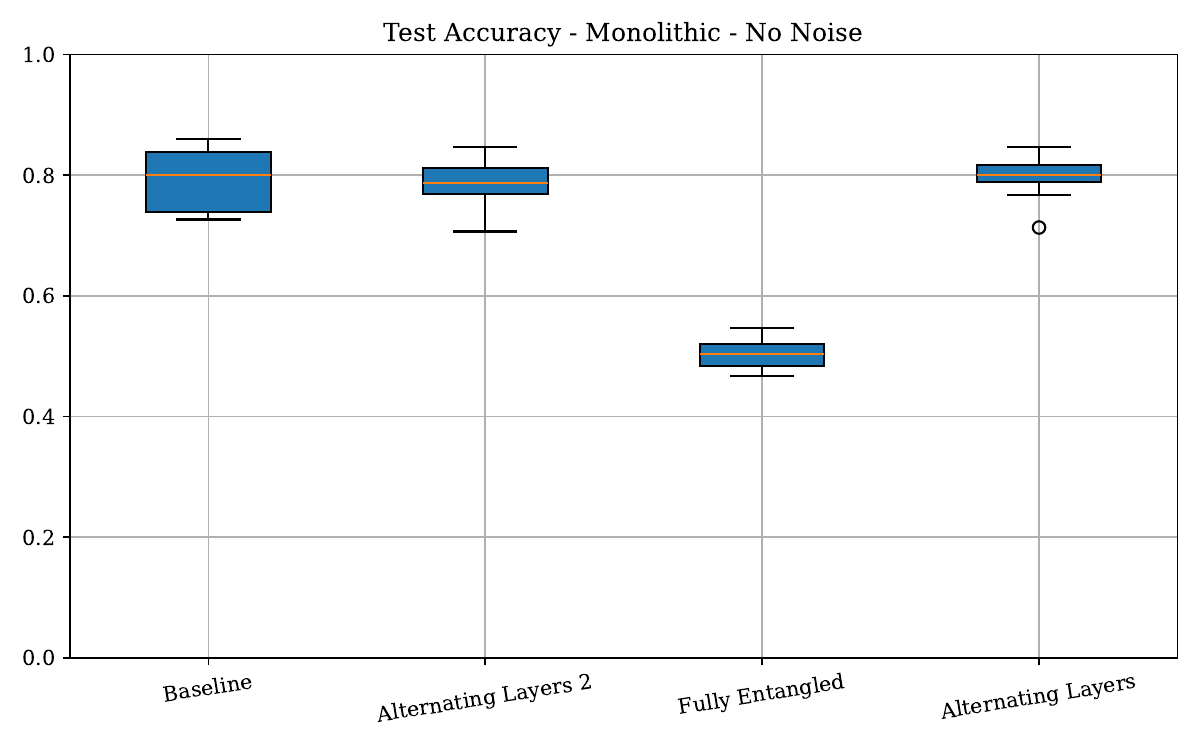}
        \caption{Test accuracy for dataset 2.}
        \label{fig:test_acc_dataset2_monolithic_no_noise}
    \end{subfigure}
    \hfill
    \begin{subfigure}{1\textwidth}
        \centering
        \includegraphics[scale=0.4]{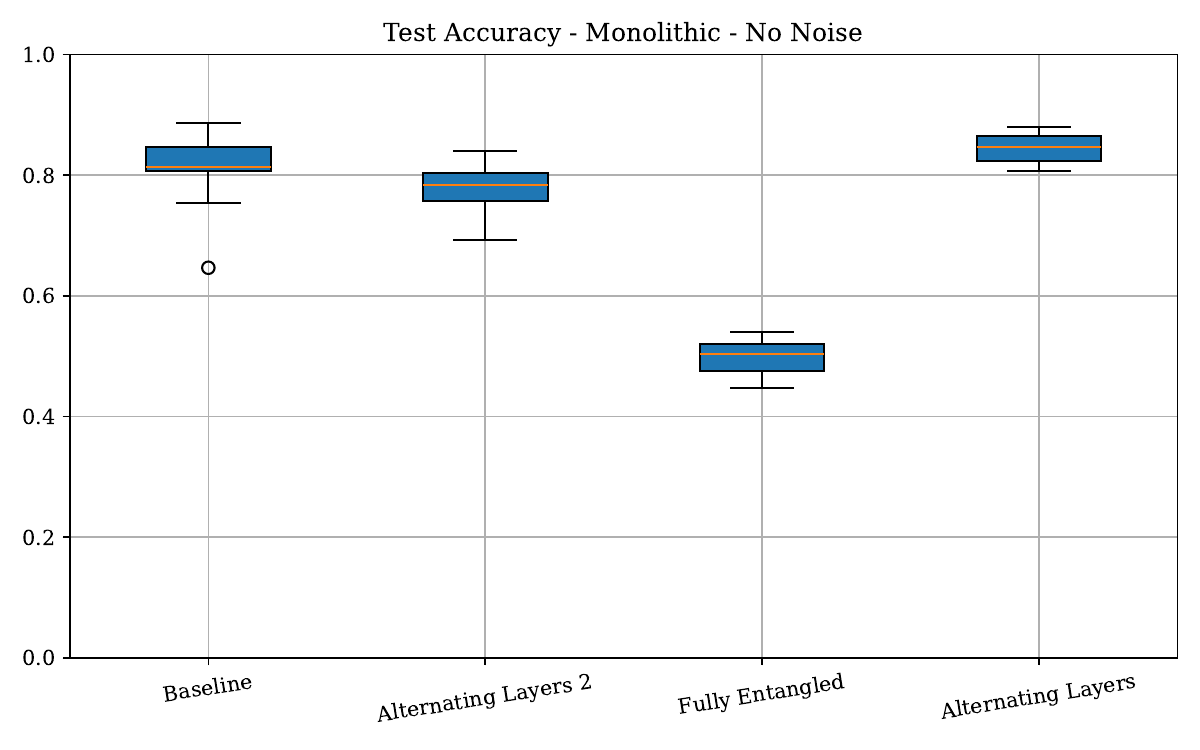}
        \caption{Test accuracy for dataset 3.}
        \label{fig:test_acc_dataset3_monolithic_no_noise}
    \end{subfigure}
    \caption{Results of the monolithic (i.e., non-distributed) circuits under ideal conditions. }
    \label{fig:results_monolithic_no_noise}
\end{figure*}

\section{Experiments}\label{sec:results}
We will start this section by discussing the experimental setup and parameters for all experiments, before presenting the results.

\subsection{Experimental Setup}\label{sec:experimental_setup}
The circuits were implemented using Qiskit \cite{qiskit2024} and are trained using the SPSA optimizer (also from Qiskit) with the objective in Eq. \ref{eqc:cost} with the log loss, i.e., cross entropy loss function implemented by Scikit-learn \cite{scikit-learn}. We evaluated the models on three different binary classification datasets created with Scikit-learn. The experiments were run for ten seeds and the configuration of the hyperparameters is shown in Table. \ref{tab:experiment_config}. The noise model used depolarizing error on all qubits with a probability of 0.03, which is only used during testing, that is, the circuits are trained under optimal conditions. Recall that the circuits are trained in a monolithic fashion, i.e., not distributed. For testing, the remote-CX operations and communication qubits are added as described in the section above. However, for comparison, we also ran the monolithic versions on the test set under ideal (i.e., without noise) conditions; these results are shown in Fig. \ref{fig:results_monolithic_no_noise}.

\begin{table}[tb]
    \centering
    \caption{Experimental configuration}
    \begin{tabular}{lcc}
    \toprule
         VQC qubits & 8 \\ \midrule
         QPU capacity (computational) & 2 \\ \midrule
         QPU capacity (comm.) & 2 \\ \midrule
         QPUs & 4 \\ \midrule
         VQC Layers & 10 \\ \midrule
         Shots & 1000 \\ \midrule
         %Features & 8 \\ \midrule
         Iterations & 1000 \\ \midrule
         Dataset size & 1000 \\ \midrule
         Train/Validation/Test split & 700/150/150 \\   \midrule
         Samples per iteration & 64 \\
    \bottomrule
    \end{tabular}
    \label{tab:experiment_config}
\end{table}

\subsection{Results}
The results of the experiments on the first dataset are shown in Fig. \ref{fig:validation_acc_dataset_1} and Fig. \ref{fig:test_acc_dataset1}. The best performance is achieved by the ''Alternating Layers'' circuit closely followed by the baseline circuit with both achieving an accuracy close to 0.9. The ''Alternating Layers 2'' circuit has a mean accuracy of around 0.85 and ''fully entangled'' circuit only slightly above 0.6. The test accuracy of this dataset is shown in Fig. \ref{fig:test_acc_dataset1}. The best median performance is achieved by the ''Alternating Layers'' circuit while test accuracy of the ''Alternating Layers 2'' circuit is slightly above the baseline VQC. The ''fully entangled'' circuit achieves a median test accuracy of just below 0.6. 
Results for the second dataset are shown in Fig. \ref{fig:validation_acc_dataset_2} and Fig. \ref{fig:test_acc_dataset2}. The baseline as well as the ''Alternating Layers'' circuits achieve similar accuracy on the validation set followed by ''Alternating Layers 2''. The ''fully entangled'' circuit is not able to learn at all on this dataset as it remains at 0.5 accuracy throughout training. The ''Alternating Layers'' circuit is more robust against noise as can be seen in the test results when compared to the baseline and ''Alternating Layers 2'' circuits.
Performance on the third dataset is shown in Fig. \ref{fig:validation_acc_dataset_3} and Fig. \ref{fig:test_acc_dataset3}. The best performance during training is achieved by the ''Alternating Layers'' circuit with around 0.8 accuracy on the validation set. The baseline and the ''fully entangled'' circuit perform almost the same and achieve an accuracy slightly below 0.8. The ''fully entangled'' circuit again fails to learn and remains at a mean accuracy of 0.5. The ''Alternating Layers'' circuit again delivers the best performance on the test set and demonstrates its superior robustness towards noise compared to the other circuits. The baseline and ''Alternating Layers 2'' circuits have almost identical median test accuracy, with the latter having overall more spread out results.

\section{Discussion}
Overall, the baseline, ''Alternating Layers'', and the ''Alternating Layers 2'' circuits perform fairly similar during training, with the former two achieving almost identical accuracy on the validation set in two experiments. These circuits have different entangling patterns in each layer; the ''fully entangled'' circuit, however, which entangles all qubits at the beginning of the circuit and contains no further entanglement, fails to achieve an improvement in two of the three experiments and only a small increase in accuracy during training in the third. Thus, in these experiments, a form of entanglement at least in some of the layers was required to allow for meaningful improvement.
When the circuits are transformed to their DQC equivalent version and executed under noise, the ''Alternating Layers'' circuit, which has less remote operations than the baseline, is more robust and achieves the best performance in all experiments. This was expected as less remote operations result in a more compact circuit that is subject to less noise. However, the experiments have also shown that simply removing CX gates is not sufficient. For instance, the circuit in which all qubits are entangled in the beginning and have no further entangling gates in each layer was unable to learn at all. Moreover, the ''Alternating Layers 2'' circuit which also has less remote operations than the baseline VQC as well as the ''Alternating Layers'' circuit is not able to achieve the same training performance as the baseline, though it is more stable in the DQC setting under noise, that is, its performance does not drop as much as the baseline. The results indicate that the entangling patterns used in the circuit architecture, i.e., what qubits are entangled at what point in the circuit, are both generally important during training in a monolithic setting, however, they become even more important in the distributed paradigm. With a higher noise probability or more sophisticated noise models the discrepancy may increase even further. 
While this study highlights the importance of selecting appropriate entanglement between qubits, especially when remote gates must be implemented, it is crucial to point out some limitations. As mentioned above, the implemented framework is not a full quantum network simulator, and protocols such as entanglement swapping or purification were not applied. The fidelity of entanglement and entanglement generation required for communication was also not considered. Instead, the focus was on the VQC architectures and how the remote-CX protocol would affect the performance when the trained models are executed under noise. That is, we transformed the circuits to a logical equivalent setting of 4 QPUs each with 2 data and communication qubits. The protocols to perform a remote CX were then inserted at the corresponding non-local gates. Thus, in real-world or more realistic network settings, the effect of unnecessary remote operations may be more pronounced or noticeable. Simulating and training VQCs with more than a few qubits on a network simulator is currently still challenging and computationally expensive, however, should be investigated in future work. The proposed approach in this paper provides a simplified framework in which the effect of remote operations and different VQC architectures can be evaluated.

\section{Conclusion}\label{sec:conclusion}
In this paper, we evaluated various VQC architectures in context of the DQC domain. This included experiments in a setting comprising 4 QPUs each with two logical and two communication qubits. Qubits were assigned to QPUs, and CX gates between QPUs were implemented using the remote-CX protocol. Circuits were trained for a binary classification task under ideal conditions in a non-distributed fashion and only converted to an equivalent circuit fit for the DQC paradigm for testing and evaluated under noise. All VQC architectures had an equal number of trainable parameters; the difference between the circuits was how often and between what qubits entangling gates were applied. The results indicated that the entanglement patterns can have a significant influence on the performance, especially compared to a circuit architecture that only creates a fully entangled state at the start of the circuit. However, simply adding or removing entangling gates may not be sufficient to improve performance in both training and testing. Rather, VQC architectures that achieve a balance between local and remote operations are likely to be more suitable for the DQC paradigm. 

\bibliographystyle{IEEEtran}
\bibliography{IEEEabrv,bibliography}

\begin{thebibliography}{10}
\providecommand{\url}[1]{#1}
\csname url@samestyle\endcsname
\providecommand{\newblock}{\relax}
\providecommand{\bibinfo}[2]{#2}
\providecommand{\BIBentrySTDinterwordspacing}{\spaceskip=0pt\relax}
\providecommand{\BIBentryALTinterwordstretchfactor}{4}
\providecommand{\BIBentryALTinterwordspacing}{\spaceskip=\fontdimen2\font plus
\BIBentryALTinterwordstretchfactor\fontdimen3\font minus \fontdimen4\font\relax}
\providecommand{\BIBforeignlanguage}[2]{{%
\expandafter\ifx\csname l@#1\endcsname\relax
\typeout{** WARNING: IEEEtran.bst: No hyphenation pattern has been}%
\typeout{** loaded for the language `#1'. Using the pattern for}%
\typeout{** the default language instead.}%
\else
\language=\csname l@#1\endcsname
\fi
#2}}
\providecommand{\BIBdecl}{\relax}
\BIBdecl

\bibitem{schuld2015introduction}
M.~Schuld, I.~Sinayskiy, and F.~Petruccione, ``An introduction to quantum machine learning,'' \emph{Contemporary Physics}, vol.~56, no.~2, pp. 172--185, 2015.

\bibitem{biamonte2017quantum}
J.~Biamonte, P.~Wittek, N.~Pancotti, P.~Rebentrost, N.~Wiebe, and S.~Lloyd, ``Quantum machine learning,'' \emph{Nature}, vol. 549, no. 7671, pp. 195--202, 2017.

\bibitem{cerezo2022challenges}
M.~Cerezo, G.~Verdon, H.-Y. Huang, L.~Cincio, and P.~J. Coles, ``Challenges and opportunities in quantum machine learning,'' \emph{Nature computational science}, vol.~2, no.~9, pp. 567--576, 2022.

\bibitem{maheshwari2022quantum}
D.~Maheshwari, B.~Garcia-Zapirain, and D.~Sierra-Sosa, ``Quantum machine learning applications in the biomedical domain: A systematic review,'' \emph{Ieee Access}, vol.~10, pp. 80\,463--80\,484, 2022.

\bibitem{wei2023quantum}
L.~Wei, H.~Liu, J.~Xu, L.~Shi, Z.~Shan, B.~Zhao, and Y.~Gao, ``Quantum machine learning in medical image analysis: A survey,'' \emph{Neurocomputing}, vol. 525, pp. 42--53, 2023.

\bibitem{ullah2024quantum}
U.~Ullah and B.~Garcia-Zapirain, ``Quantum machine learning revolution in healthcare: a systematic review of emerging perspectives and applications,'' \emph{IEEE Access}, vol.~12, pp. 11\,423--11\,450, 2024.

\bibitem{orus2019quantum}
R.~Or{\'u}s, S.~Mugel, and E.~Lizaso, ``Quantum computing for finance: Overview and prospects,'' \emph{Reviews in Physics}, vol.~4, p. 100028, 2019.

\bibitem{pistoia2021quantum}
M.~Pistoia, S.~F. Ahmad, A.~Ajagekar, A.~Buts, S.~Chakrabarti, D.~Herman, S.~Hu, A.~Jena, P.~Minssen, P.~Niroula \emph{et~al.}, ``Quantum machine learning for finance iccad special session paper,'' in \emph{2021 IEEE/ACM international conference on computer aided design (ICCAD)}.\hskip 1em plus 0.5em minus 0.4em\relax IEEE, 2021, pp. 1--9.

\bibitem{preskill2018quantum}
J.~Preskill, ``Quantum computing in the nisq era and beyond,'' \emph{Quantum}, vol.~2, p.~79, 2018.

\bibitem{kimble2008quantum}
H.~J. Kimble, ``The quantum internet,'' \emph{Nature}, vol. 453, no. 7198, pp. 1023--1030, 2008.

\bibitem{caleffi2024distributed}
M.~Caleffi, M.~Amoretti, D.~Ferrari, J.~Illiano, A.~Manzalini, and A.~S. Cacciapuoti, ``Distributed quantum computing: a survey,'' \emph{Computer Networks}, vol. 254, p. 110672, 2024.

\bibitem{cacciapuoti2019quantum}
A.~S. Cacciapuoti, M.~Caleffi, F.~Tafuri, F.~S. Cataliotti, S.~Gherardini, and G.~Bianchi, ``Quantum internet: Networking challenges in distributed quantum computing,'' \emph{IEEE Network}, vol.~34, no.~1, pp. 137--143, 2019.

\bibitem{briegel1998quantum}
H.-J. Briegel, W.~D{\"u}r, J.~I. Cirac, and P.~Zoller, ``Quantum repeaters: the role of imperfect local operations in quantum communication,'' \emph{Physical Review Letters}, vol.~81, no.~26, p. 5932, 1998.

\bibitem{van2013designing}
R.~Van~Meter and J.~Touch, ``Designing quantum repeater networks,'' \emph{IEEE Communications Magazine}, vol.~51, no.~8, pp. 64--71, 2013.

\bibitem{azuma2023quantum}
K.~Azuma, S.~E. Economou, D.~Elkouss, P.~Hilaire, L.~Jiang, H.-K. Lo, and I.~Tzitrin, ``Quantum repeaters: From quantum networks to the quantum internet,'' \emph{Reviews of Modern Physics}, vol.~95, no.~4, p. 045006, 2023.

\bibitem{ferrari2021compiler}
D.~Ferrari, A.~S. Cacciapuoti, M.~Amoretti, and M.~Caleffi, ``Compiler design for distributed quantum computing,'' \emph{IEEE Transactions on Quantum Engineering}, vol.~2, pp. 1--20, 2021.

\bibitem{cerezo2021variational}
M.~Cerezo, A.~Arrasmith, R.~Babbush, S.~C. Benjamin, S.~Endo, K.~Fujii, J.~R. McClean, K.~Mitarai, X.~Yuan, L.~Cincio \emph{et~al.}, ``Variational quantum algorithms,'' \emph{Nature Reviews Physics}, vol.~3, no.~9, pp. 625--644, 2021.

\bibitem{mitarai2018quantum}
K.~Mitarai, M.~Negoro, M.~Kitagawa, and K.~Fujii, ``Quantum circuit learning,'' \emph{Physical Review A}, vol.~98, no.~3, p. 032309, 2018.

\bibitem{benedetti2019parameterized}
M.~Benedetti, E.~Lloyd, S.~Sack, and M.~Fiorentini, ``Parameterized quantum circuits as machine learning models,'' \emph{Quantum science and technology}, vol.~4, no.~4, p. 043001, 2019.

\bibitem{diadamo2021distributed}
S.~DiAdamo, M.~Ghibaudi, and J.~Cruise, ``Distributed quantum computing and network control for accelerated vqe,'' \emph{IEEE Transactions on Quantum Engineering}, vol.~2, pp. 1--21, 2021.

\bibitem{khait2023variational}
I.~Khait, E.~Tham, D.~Segal, and A.~Brodutch, ``Variational quantum eigensolvers in the era of distributed quantum computers,'' \emph{Physical Review A}, vol. 108, no.~5, p. L050401, 2023.

\bibitem{du2022distributed}
Y.~Du, Y.~Qian, X.~Wu, and D.~Tao, ``A distributed learning scheme for variational quantum algorithms,'' \emph{IEEE Transactions on Quantum Engineering}, vol.~3, pp. 1--16, 2022.

\bibitem{jones2024distributed}
G.~M. Jones and H.-A. Jacobsen, ``Distributed quantum computing for chemical applications,'' in \emph{2024 IEEE International Conference on Quantum Computing and Engineering (QCE)}, vol.~2.\hskip 1em plus 0.5em minus 0.4em\relax IEEE, 2024, pp. 155--160.

\bibitem{pira2023invitation}
L.~Pira and C.~Ferrie, ``An invitation to distributed quantum neural networks,'' \emph{Quantum Machine Intelligence}, vol.~5, no.~2, p.~23, 2023.

\bibitem{neumann2022distributed}
N.~M. Neumann and R.~S. Wezeman, ``Distributed quantum machine learning,'' in \emph{International Conference on Innovations for Community Services}.\hskip 1em plus 0.5em minus 0.4em\relax Springer, 2022, pp. 281--293.

\bibitem{parekh2021quantum}
R.~Parekh, A.~Ricciardi, A.~Darwish, and S.~DiAdamo, ``Quantum algorithms and simulation for parallel and distributed quantum computing,'' in \emph{2021 IEEE/ACM Second International Workshop on Quantum Computing Software (QCS)}.\hskip 1em plus 0.5em minus 0.4em\relax IEEE, 2021, pp. 9--19.

\bibitem{schuld2020circuit}
M.~Schuld, A.~Bocharov, K.~M. Svore, and N.~Wiebe, ``Circuit-centric quantum classifiers,'' \emph{Physical Review A}, vol. 101, no.~3, p. 032308, 2020.

\bibitem{sim2019expressibility}
S.~Sim, P.~D. Johnson, and A.~Aspuru-Guzik, ``Expressibility and entangling capability of parameterized quantum circuits for hybrid quantum-classical algorithms,'' \emph{Advanced Quantum Technologies}, vol.~2, no.~12, p. 1900070, 2019.

\bibitem{hubregtsen2021evaluation}
T.~Hubregtsen, J.~Pichlmeier, P.~Stecher, and K.~Bertels, ``Evaluation of parameterized quantum circuits: on the relation between classification accuracy, expressibility, and entangling capability,'' \emph{Quantum Machine Intelligence}, vol.~3, no.~1, p.~9, 2021.

\bibitem{rohe2024questionable}
T.~Rohe, D.~Schuman, J.~N{\"u}blein, L.~S{\"u}nkel, J.~Stein, and C.~Linnhoff-Popien, ``The questionable influence of entanglement in quantum optimisation algorithms,'' in \emph{2024 IEEE International Conference on Quantum Computing and Engineering (QCE)}, vol.~1.\hskip 1em plus 0.5em minus 0.4em\relax IEEE, 2024, pp. 1497--1503.

\bibitem{du2022quantum}
Y.~Du, T.~Huang, S.~You, M.-H. Hsieh, and D.~Tao, ``Quantum circuit architecture search for variational quantum algorithms,'' \emph{npj Quantum Information}, vol.~8, no.~1, p.~62, 2022.

\bibitem{ding2022evolutionary}
L.~Ding and L.~Spector, ``Evolutionary quantum architecture search for parametrized quantum circuits,'' in \emph{Proceedings of the genetic and evolutionary computation conference companion}, 2022, pp. 2190--2195.

\bibitem{situ2024distributed}
H.~Situ, Z.~He, S.~Zheng, and L.~Li, ``Distributed quantum architecture search,'' \emph{Physical Review A}, vol. 110, no.~2, p. 022403, 2024.

\bibitem{qiskit2024}
A.~Javadi-Abhari, M.~Treinish, K.~Krsulich, C.~J. Wood, J.~Lishman, J.~Gacon, S.~Martiel, P.~D. Nation, L.~S. Bishop, A.~W. Cross, B.~R. Johnson, and J.~M. Gambetta, ``Quantum computing with {Q}iskit,'' 2024.

\bibitem{scikit-learn}
F.~Pedregosa, G.~Varoquaux, A.~Gramfort, V.~Michel, B.~Thirion, O.~Grisel, M.~Blondel, P.~Prettenhofer, R.~Weiss, V.~Dubourg, J.~Vanderplas, A.~Passos, D.~Cournapeau, M.~Brucher, M.~Perrot, and E.~Duchesnay, ``Scikit-learn: Machine learning in {P}ython,'' \emph{Journal of Machine Learning Research}, vol.~12, pp. 2825--2830, 2011.

\end{thebibliography}

\end{document}